\documentclass[aps,prb,superscriptaddress]{revtex4-1}

\usepackage[dvips]{graphicx}
\usepackage{longtable}
\usepackage{multirow}
\usepackage{amsmath}
\usepackage{color}

\begin{document}

\preprint{}

\title{
Atomistic structure of amorphous silicon nitride from classical molecular dynamics simulations.
}


\author{Mariella Ippolito}
\affiliation{
Consorzio Interuniversitario per le Applicazioni di Supercalcolo
Per Universit\`a e Ricerca (CASPUR),\\
Via dei Tizii 6, 00185 Roma, Italy
}

\author{Simone Meloni}
\email{To whom the correspondence should be addressed: simone.meloni@caspur.it}
\affiliation{
School of Physics, Room 126 UCD-EMSC, University College Dublin, 
Belfield, Dublin 4, Ireland 
}


\date{\today}

\begin{abstract}

By means of molecular dynamics simulations based on the Billeter et al. [S. R. Billeter, A. Curioni, D. Fischer, and W. Andreoni, Phys. Rev. B {\bf 73}, 155329] environment-dependent classical force field we studied the structural features of SiN$_x$ samples at various stoichiometries. Our results are in good agreement with experimental data and are able to reproduce some features which so far were not reproduced by simulations. In particular, we identified units containing N--N bonds, which are thought to be responsible for an unassigned peak in the radial distribution function obtained from neutron diffraction data and signals observed in electron spin resonance, X-ray photoemission spectroscopy, electron-energy-loss spectroscopy and optical absorption experiments. 

We have identified defects which are thought to be the responsible for the high concentration of charge traps that makes this material suitable for building non-volatile memory devices. We analyzed the dependency of the concentration of these defects with the stoichiometry of the sample.

\end{abstract}


\maketitle

\section{Introduction
\label{Introduction}}
Amorphous Silicon Nitride (a-Si$_3$N$_4$, or a-SiN$x$ if the alloy is non-stoichiometric) is a very promising materials for application in the field of memory devices \cite{bachhofer:2791} and optoelectronics\cite{kim:123102,ma:2006,deshpande:6534, deshpande:6534,warren:3685,Liu20042397}.
As for optoelectronics, in Si rich a-SiN$_x$ samples Si clusters (Si-c) present in the matrix \cite{negro:183103} are thought to be responsible for the luminescence of this material.\cite{kim:123102,ma:2006} However, a-SiN$_x$ films are known to be luminescent matrices on their own due to the light emission related to the presence of specific defects.\cite{deshpande:6534, deshpande:6534,warren:3685,Liu20042397} 
Si and N related defects act also as charge traps and their high concentration in a-SiN$_x$ makes this material suitable for building non-volatile memory devices.

Driven by its applicative importance, a significant effort has been made to clarify the atomistic and electronic structure of SiN$_x$, and the relation between them.\cite{jncs-si3n4-xray,jncs-si3n4,jjap.32.876, PhysRevB.42.5677, lin-optoAndAdvMat-4_543,lin-optmat-93-93,PhysRevB.65.073202,prb80-144201,prb82-205201,prb-68_205203,PhysRevB.58.8323}
On the one hand, the atomistic structure has been investigated  by X-ray\cite{jncs-si3n4-xray} and neutron\cite{jncs-si3n4} diffraction experiments. These experiments revealed a significant degree of under-coordination of Si and N atoms, reporting and average coordination of 3.7 and 2.78 for Si and N, respectively. Another interesting structural feature found in the neutron Radial Distribution Function (RDF) is a peak at about $1.3$~\AA. Such a short distance is only compatible with the presence of covalent bonds between N atoms in the sample. These bonds cannot be due to N$_2$ molecules trapped in the sample as in this molecule the bond length is shorter ($1.1$~\AA). An hypothesis is that this peak is due to units in which one N atom is bonded to another N atom and to Si atoms. As only the total RDF was measured, and the atomistic model used for its interpretation did not include N--N bonds, this hypothesis cannot be confirmed. As a result, the structure of SiN$_x$ reported in Refs. \onlinecite{jncs-si3n4-xray} and \onlinecite{jncs-si3n4} looks, locally, very similar to the crystal one. 
The presence of N--N bonds in hydrogenated and non-hydrogenated SiN$_x$ samples is further supported by Electron Spin Resonance (ESR)\cite{jjap.32.876}, X-ray photoemission spectroscopy, electron-energy-loss spectroscopy and optical absorption\cite{PhysRevB.42.5677} experiments. 

Several computational studies have been performed to identify the atomistic structure and to compute the corresponding electronic structure of SiN$_x$. Some of them used very rough atomistic models, such as the Bethe lattice\cite{lin-optoAndAdvMat-4_543,lin-optmat-93-93}, in which the atomistic structure is decided {\em a priori}. We shall not compare our results with these works as their goal is to determine the dependence of the electronic structure from the atomistic configuration, without considering the statistical relevance of the (predetermined) configuration used in the calculation. In others works, the a-SiN$_x$ samples were generated by using procedures based on the randomization of either the atomic positions or the bonds\cite{kroll2001,prb-68_205203}, or on the combination of randomly oriented subunits\cite{PhysRevB.54.R15594}. Sometimes the random-orientation step was followed by a relaxation step performed by low-temperature Molecular Dynamics (MD) or Monte Carlo (MC) runs. 
Finally, in some studies \cite{PhysRevB.41.12197,PhysRevB.65.073202,PhysRevB.58.8323,prb80-144201,prb82-205201} the samples were prepared by the quench-from-the-melt procedure (see Sec. \ref{sec:ComputationalDetails}). While all these studies are able to reproduce the general features of the experimental RDF, in none of them it is reported the formation of N--N bonds. 
Only the pair distribution function reported in Ref.~[\onlinecite{prb-68_205203}] shows indeed a small peak at $\sim 1.3$~\AA, but this feature is not discussed by the authors. Moreover, its intensity is very low compared with experimental data.\cite{jncs-si3n4}
The failure of classical and, especially, {\em ab initio} force models to reproduce this covalent bond is surprising. We believe that the reason of this failure has to be sought in the procedure by which the samples were generated rather than in the accuracy of the force models used for the simulations, with the exception of Ref.s~[\onlinecite{PhysRevB.65.073202}] and [\onlinecite{PhysRevB.58.8323}] in which it is used a force model that explicitly prevents the formation of N--N. 
In fact, in the samples produced by random orientation of subunits the formation of the N--N bonds is prevented by the procedure that imposes the conservation of the Si\slash N alternation. As for the samples prepared by {\it ab initio} quenching-from-the-melt procedures\cite{prb80-144201,prb82-205201}, the melting temperature is too low ($3000-3500$~K) and the melting and cooling times are too short (few picoseconds each). In these conditions the breaking of Si--N bonds and the formation of N--N bonds, required for the formation of units involving N--N bonds, cannot take place. In fact, for a similar material (a-SiO$_2$) Vollmayr et al. \cite{PhysRevB.54.15808} have demonstrated that higher temperatures ($7000$~K) and very long melting and cooling processes (up to $100$~ps and $1$~ns, respectively) are needed to obtain good amorphous samples. A special consideration is in order for Refs.~[\onlinecite{PhysRevB.58.8323}]. In fact, the results reported in this paper are obtained using a potential in which the N--N interactions are purely repulsive and therefore the formation of N--N bonds is prevented.  Also Vashishta and Kalia\cite{prl75-858} did not find N--N bonds in their samples generated via classical MD simulations. In this case the melting temperature, the melting time and the cooling rate are adequate. 
In this case, we believe that the inability to produce units containing the N--N bond is probably due to a limitation of the force field adopted for the simulations.

The methods and the conditions used for generating the samples could have introduced also other artifacts in the corresponding atomistic structures. For example, they could have produced samples with a too low concentration of Si and N dangling bonds, over-coordinated Si and Si--Si bonds, which are all thought to be defects influencing the electronic and optical properties of the material. The aim of this work is therefore to further investigate the atomistic structure of a-SiN$_x$ at various stoichiometries ($0.3 \leq x \leq 1.8$) using the quenching-from-the-melt procedure. We shall extensively explore the domain of the parameters governing this procedure: i) melting temperature, ii) duration of the melting and iii) quenching rate. In addition to a general agreement with the experimental RDF, we shall use the formation of N--N bonds, and correspondingly the agreement with neutron RDF at $\sim 1.3$~\AA, as a benchmark for assessing the quality of the samples obtained. We shall further validate these structures by performing {\itshape ab initio} geometry optimization and MD simulations.
On good samples we shall also identify the other relevant structural parameters, such as concentration of coordination defects, and concentration and structural parameters of the N(Si$_m$N$_n$) and Si(Si$_m$N$_n$) subunits contained in them.

In our study we perform classical MD simulations based on the environment dependent force field  recently developed Billeter et al.\cite{PhysRevB.73.155329, PhysRevB.79.169904}. The ability of this force field to model the equilibrium structure and the dynamical properties of the closely related a-SiO$_2$ system has already been established\cite{fischer:012101,PhysRevB.81.014203,ApplPhysLett.93.153109}. This makes us confident that this potential is also able to properly represent the interatomic interactions in SiN$_x$ as well. However, to the best of our knowledge, no works have been published before which make use of this potential to study a-SiN$_x$. Therefore, another objective of this research is the validation of this potential also for this material.

The paper is organized as follows: in Sec. \ref{sec:ComputationalDetails} we give the computational details of our simulations. In Sec. \ref{sec:ResultsAndDiscussion} we shall present the results of our simulations and the analysis of the atomistic structure of the a-SiN$_x$ samples. Finally, in Sec. \ref{sec:Conclusions}
we shall draw some conclusions.

\section{Computational details
\label{sec:ComputationalDetails}
}
In this paper we generate and analyze the structure of amorphous SiN$_x$ sample at various compositions as obtained from MD simulations based on the Billeter et al. environment dependent force field.\cite{PhysRevB.73.155329, PhysRevB.79.169904} This force field allows to treat silicon oxide and oxynitride systems, both hydrogenated and non hydrogenated. The Billeter et al. force field is an evolution of the Tersoff potential\cite{PhysRevB.37.6991} and one of its advantages is that it describes more accurately coordination defects. In fact, one problem with the original Tersoff potential was the tendency to produce structures with an high abundance of coordination defects. In the Billeter et al. potential this problem is solved by adding a ``penalty'' term that depends on  the deviation from the optimal coordination. \textcolor{black}{Moreover, with specific reference to the case of aSiN$_x$ and at a variance form previous classical force field (e.g. the de Brito Mota et al. force field\cite{prb-68_205203}]), the Billeter et al. potential takes explicitly into account the formation of N--N bonds.} The reliability of this force field to model complex systems and phenomena, such as Si/SiO$_2$ interface\cite{ApplPhysLett.88.012101}, Si nanoparticles embedded in amorphous SiO$_2$\cite{ApplPhysLett.93.153109}, Si and O self-diffusion in SiO$_2$\cite{PhysRevB.81.014203} and strain distribution in Si nanowires\cite{tuma:193106} has been already proven. On the contrary, to the best of our knowledge, a careful validation on realistic silicon nitride systems is still lacking. This validation will therefore be another objective of this work. 

The a-Si$_3$N$_4$ sample was obtained by means of the quenching-from-the-melt method. This technique consists in melting a crystal of proper stoichiometry at high temperature and then slowly cooling it down to the target temperature, at which all the relevant properties are computed. The cooling step of the process can be performed in different ways, for example by a constant temperature MD in which the target temperature is continuously decreased or by running a series of consecutive constant temperature MD simulations in which the target temperature is decreased when passing from a run to the next. In this paper we have used the first approach (continuous cooling). However, we have also prepared a selected set of samples following the stepwise cooling procedure. The comparison of the structural characteristic of the samples prepared with the two methods has shown that there is no significant difference between them provided that the average cooling rate (i.e. $dT / dt$) is the same. Summarizing, the parameters controlling the quenching-from-the-melt procedure are: the melting temperature $T_m$, the duration of the melting step $t_m$ and the quenching rate $dT / dt$. The effect of these parameters on the structural properties of the amorphous system has been evaluated by producing several samples according to the values reported in Table \ref{tab:amorphizationProtocol}. These results are described in detail in the next section. 

The initial configuration for the quenching-from-the-melt procedure for the stoichiometric sample was the crystal structure of the $\alpha$-Si$_3$N$_4$ phase.
The density was kept fix to the experimental density of the a-Si$_3$N$_4$ ($\rho = 3.14$~g/cm$^3$\cite{PhysRevB.42.5677}), which is slightly lower than the density of $\alpha$-Si$_3$N$_4$ ($\rho \sim 3.2$g/cm$^3$).
In order to check for finite size effects, we ran simulations on samples containing as many as $224$, $1792$ and $14336$ atoms without observing any significant difference between the structure (RDF, angular distribution function and coordination defects) of the samples. This means that there is no correlation beyond the maximum length-scale compatible with the simulation box of the 224 atoms sample, that is $\sim 5.8$\AA. This is indeed in agreement with the observation that the g(r) of the amorphous samples do not show any correlation beyond $3.5$~\AA\ (see next section). All the results reported in Sec. \ref{sec:ResultsAndDiscussion} refer to the smallest sample.

The amorphous sub(supra)-stoichiometric samples were obtained starting from the a-Si$_3$N$_4$ sample and randomly replacing N (Si) atoms with Si (N) atoms. In particular, we generated SiN$_x$ samples at $x=0.4, 0.6, 0.8, 1.0, 1.2, 1.4, 1.8$. The simulation box was isotropically modified so as to obtain a density consistent with the experimental density,\cite{PhysRevB.42.5677} which varies with the composition (see  Table \ref{tab:density}). The so obtained samples are melt at $T_m = 6000$~K for $t_m = 100$~ps and then quenched down to room temperature with a quenching rate of $dT / dt = 5$~K/ps, that is at the lowest quenching rate used for generating the stoichiometric sample.

\textcolor{black}{We further checked the reliability of our results by comparing the theoretical density of all the samples at ambient conditions with experimental data.\cite{PhysRevB.42.5677} To this end, we performed  constant pressure MD simulations according to the Martyna-Tobias-Klein method\cite{constantPressure} for non-isotropic cells starting from the equilibrated samples at fixed volume (see above).}

\textcolor{black}{Finally, on samples at composition $x = 0.4, 1.333$ and $1.8$, i.e. in the sub-stoichiometrical, stoichiometrical and supra-stoichiometrical domains, we performed {\itshape ab-initio} geometry optimization and MD simulations. These calculations have been performed using a setup as close as possible to the one used by Billeter et al. to generate the training set used for fitting the parameters of their force field. In particular, the calculations have been performed using the version $3.13.2$ of the CPMD simulation package. \cite{cpmd} The electronic structure calculations were performed using the Density Functional Theory in the generalized gradient approximation of Perdew-Burke-Ernzerhof.\cite{PBE} The Kohn-Sham orbitals were expanded on a plane wave basis set with a $80$~Ry cutoff. Finally, electron-ion interactions are described using Troullier-Martins pseudopotentials.\cite{TM}}

\section{Results and discussion
\label{sec:ResultsAndDiscussion}}
The general aspects of the quenching-from-the-melt amorphization protocol were presented in sec. \ref{sec:ComputationalDetails}. The ability of the protocol to produce a realistic amorphous sample depends on the parameters controlling the various steps. In particular, the parameters must be chosen such that in the first step the sample is completely melt and in the second step the sample can relax to a local minimum of the free energy. 
In this work we have ran a series of simulations at various $T_m$, $t_m$ and $dT/dt$. Moreover, we have evaluated the effect of the stoichiometry on the structure of the sample by studying, in addition to the stoichiometric Si$_3$N$_4$ sample, sub and supra stoichiometric SiN$_x$ samples, with $x=0.4, 0.6, 0.8, 1, 1.2, 1.4, 1.8$. 
In table \ref{tab:amorphizationProtocol} we summarize the list of simulations performed on the Si$_3$N$_4$ sample and the corresponding values of the parameters governing the quench-from-the-melt procedure.

Before presenting and discussing our results, it is worth to comment on the melting temperatures used in our simulations. The experimental melting temperature of the Si$_3$N$_4$ depends on the conditions under which the sample is prepared and the melting temperature is measured. Commercial samples (e.g. a-Si$_3$N$_4$ from Atlantic Equipment Engineers, Div. of Micron Metals, Inc. - Bergenfield, NJ) are reported to melt at about 2175~K (other sources report a-Si$_3$N$_4$ to decompose before melting). Accurate experiments\cite{si3n4-melting} in N$_2$ atmosphere have shown the a-Si$_3$N$_4$ to melt approximately at $2775$~K. \textcolor{black}{However, in simulations the melting process typically occurs at much higher temperatures. There are several reasons for this, of which the accuracy of the force field is possibly not the most relevant one. In simulations we typically start from the perfect crystal while in real samples there is a natural abundance of defects, such as vacancies, which allow a larger atomic mobility. Moreover, especially in constant volume MD simulations of small samples, the fluctuation of local density is strongly limited. Fluctuations of the density naturally occurring in real samples enhances atomic diffusion which, in turn, produces the melting. The higher temperature needed for observing the melting in simulations might also be due to another phenomenon, namely that the formation of a liquid in a bulk crystal (i.e. in absence of a surface, as it is the case in all the simulations mentioned in the introduction and in the present work) this process is termally activated (i.e. there is a barrier in the free energy profile separating the crystal from the liquid domain). An exhaustive description of the theory of the nucleation of a liquid phase in a bulk crystal, and the relative experimental and computational results, is given in the Chapter 3 of Ref. \onlinecite{KeltonGreer}. This means that at the computational melting temperature (i.e. the temperature at which the liquid phase is more stable than the crystal phase given the specific force field) the system can stay trapped in a metastable state corresponding to the overheated crystal for the entire duration of the simulation.} Therefore, in order to obtain a liquid sample starting from a crystal on the short timescales allowed by simulations (from few $ps$ in {\it ab-initio} simulation up to several hundred $ps$ in classical MD with environment dependent classical force fields) it is necessary to go much beyond the experimental or computational melting temperature.This explains why in our simulations we perform the melting at very high temperatures (all above 5000~K). This is in agreement with what was done in analogous studies on a related material using a different force field.\cite{PhysRevB.54.15808}

In 
Fig.~1
it is reported the pair correlation function ($g(r)$) of the Si$_3$N$_4$ melt at various temperatures measured over 10~ps of constant temperature MD after initial $100$~ps of thermalization at the target temperature T$_m$. For comparison, we also report $g(r)$ of the Si$_3$N$_4$ crystal at 300~K. In the inset of 
Fig.~1 
we report the N--N partial g(r) at the same temperatures. 
The first remark is that for T$_m \leq 5500$~K the structure preserves a long range order. For example, these samples still have peaks in the g(r) at $\sim 3.4 $~\AA\ and $\sim 4.1 $~\AA\, and a series of superimposed peaks between $\sim 4.3-4.8$~\AA. Moreover, the broad peak at $\sim 2.8$~\AA\ still shows traces of the three peaks present in the crystal structure.  This means that temperatures $\leq$ 5500~K do not produce adequate melting on a timescale of 100~ps. 
In these samples, there is no evidence of the peak at $\sim 1.3$~\AA.
However, for slightly higher temperatures (T$_m \ge 5800$~K), the long range features in the g(r) disappear. This means that the sample is completely melted on the $100$~ps timescale of the simulation. In these liquid samples it also appears the peak at $1.3$~\AA. The shape of the g(r) in general, and the peak at $1.3$~\AA\ in particular, are only marginally affected by a further increase of the temperature. This indicates that the liquid samples produced by high temperature simulations have reached the equilibrium. 

We now turn to the analysis of the origin of the peak at $1.3$~\AA. Such a short distance is compatible only with N--N bonds. This is confirmed by the presence in the N--N partial g(r) of a peak at $1.3$~\AA\ (see inset in 
Fig.~1). 
This result is \textcolor{black}{consistent} with experimental data.\cite{jncs-si3n4,jjap.32.876,PhysRevB.42.5677} 
\textcolor{black}{The presence of N--N bonds indicates that in the melt the N atoms are not only part of the N(Si$_3$)/Si(N$_4$) units, which are the only units present in the crystal. Let us indicate by N(Si$_{n}$N$_m$) a generic unit in which $m$ nitrogen and $n$ silicon atoms are bonded to a ``central'' N atom, and by Si(Si$_{n}$N$_m$) a unit in which $m$ nitrogen and $n$ silicon atoms are bonded to a ``central'' Si atom. The concentration of the N(Si$_{n}$N$_m$) units can be used to monitor the stoichiometric defects of the central N atom and the concentration of the Si(Si$_{n}$N$_m$) units those of the central Si atoms. In table \ref{tab:structuralDataMelt} is reported the concentration and the structure of the more abundant of such units at $5800$~K, $6000$~K and $7000$~K. Let us start by analyzing the stoichiometric defects of the N atoms. In the melt at $5800$~K, the most abundant N(Si$_3$)/Si(N$_4$) units are (in descending order of concentration) N(Si$_3$) ($58\%$), N(Si$_2$N) ($23\%$),  N(Si$_2$) ($13\%$),  N(SiN) ($3.5\%$) and  N(SiN$_2$) ($3.5\%$). Consistently with what we found for the g(r), the temperature affect only marginally the abundance of these units. At $6000$~K we found a concentration of $48\%$, $23\%$, $14.5\%$, $4.5\%$ and $4\%$ for the units, N(Si$_3$), N(Si$_2$N),  N(Si$_2$),  N(SiN) and N(SiN$_2$) respectively. At $7000$~K the concentration of these units are (in the same order as before) $41\%$, $22\%$, $18.5\%$, $6.5\%$ and $4.5\%$. As expected, the main effect of the temperature is to increase the concentration of defective units (N(Si$_2$N),  N(Si$_2$),   N(SiN) and N(SiN$_2$)), some of which contain N--N bonds, and to decrease the concentration of N(Si$_3$), i.e. the  unit present in the crystalline Si$_3$N$_4$. 
As for the structure of these units, the Si--N bond length ranges from $1.77$~\AA\ to $1.90$~\AA, and the N--N one from $1.41$~\AA\ to $1.53$~\AA\ depending on the unit. As expected, the bond lengths is shorter in under-coordinated units ($n + m < 3$) and longer in over-coordinated ones ($n + m > 3$). On the contrary, the stoichiometry of the structure affects only marginally the structure. The situation is very similar for the Si(Si$_{n}$N$_m$) units: the concentration of defective units, some of which containing Si--Si bonds, increases with the temperature. Also in this case, the structure is only marginally affected by the temperature.} These results were confirmed by {\itshape ab initio} simulations. In table \ref{tab:structuralDataMelt} we compare the structural data and concentration of the N(Si$_{n}$N$_m$)
units as obtained from our classical and {\itshape ab initio} MD simulations. In particular, the latter were started from one configuration extracted from the classical simulations at corresponding temperatures. The comparison shows that both the concentration and the geometry of the classical and {\itshape ab initio} N(Si$_{n}$N$_m$ units are very close to each other, so confirming the reliability of the a-SiN$_3$ structure obtained from the classical simulations.

Also the quenching rate is expected to influence the structure of the a-Si$_3$N$_4$, we therefore studied its effect by performing a series of simulations at various $dT/dt$ in the range 0.5~K/ps to 500~K/ps (see Table \ref{tab:amorphizationProtocol}). For these simulations we started always from the melt generated at $T_m = 6000$~K. In 
Fig.~2 
are shown the g(r) obtained from simulations at various quenching rates. The main features of these g(r) are very similar, namely all the g(r) are characterized by three main peaks at 1.3~\AA, 1.7~\AA\ and 2.9~\AA, and another peak at 2.65~\AA\ merged with the previous one. However, there are significant differences in the details between the g(r) at the various quenching rates. The intensity of the N--N peak decreases and the intensity of the two main peaks at 1.7~\AA\ and 2.9~\AA\ increase for slower quenching rates. This dependence of the g(r) on the quenching rate can be quantified by computing the number of pairs corresponding to the different peaks. \textcolor{black}{In turn, these can be computed by properly integrating the g(r) on the intervals associated to these peaks (i.e. taking into account the Jacobian associated to the cartesian-to-polar coordinate transformation)}. However, the peak at 2.9~\AA\ is the combination of three superimposed peaks associated to second Si--Si, Si--N and N--N neighbor pairs. This peak is therefore not suitable for the analysis proposed above. For this reason we perform the analysis described above by comparing only the number of N--N and Si--N pairs corresponding to the peaks at $1.3$\AA\ and $1.7$\AA\, respectively. In the inset of 
Fig.~2 
we show the ratio of the number of Si--N pairs over the number of N--N pairs as a function of the decreasing cooling rate. This curve clearly demonstrates that this ratio increases with the decreasing of the quenching rate, eventually converging to an equilibrium value for very slow quenching rates (for $dT/dt \leq 10K/ps$). This behavior can be explained as follows. Let us consider the N(Si$_3$) and the Si(N$_4$) units, that are the building blocks composing the (stoichiometric) Si$_3$N$_4$ crystal. Assuming that these units are more stable than the units containing the N--N bond (N(Si$_n$N$_m$), see 
Fig.~3),
 which is a reasonable assumption as the second is not present in crystalline (low temperature) Si$_3$N$_4$, the relative concentration of N(Si$_n$N$_m$) units with respect to N(Si$_3$) will increase with the temperature. Therefore, at $6000$~K there will be a higher relative concentration of these kind of units with respect to the equilibrium concentration in a-Si$_3$N$_4$ at $300$~K. When the samples are quenched too quickly the system cannot relax to the equilibrium and the final relative concentration of N--N pairs is too high, i.e. it will remain close to the concentration present in the high temperature melt. On the contrary, at low quenching rates the system can relax and the N--N pairs concentration can reach its equilibrium value.
In 
Fig.~4 
we report the Radial Distribution Function (RDF) of the sample obtained at $dT/dt = 5$~K/ps, which is in very good agreement with the neutron diffraction data reported in Ref.~\onlinecite{jncs-si3n4}. \textcolor{black}{In particular, the agreement of both the position and the intensity of the computational and experimental peak at $\sim 1.3$~\AA\ is very remarkable. However, as a further validation of our results, starting from the last configuration of the classical MD simulation we performed an {\textit ab-initio} geometry optimization of the structure. As a first remark, all the N--N bonds present in the initial configuration (i.e. the one obtained from the classical MD) are still present in the final one. Moreover, the root mean square displacement (RMSD) between the {\itshape ab-initio} minimum energy configuration and the classical configuration averaged over the MD trajectory (there is no diffusion at this temperature over the timescale of the simulation and therefore this averaged configuration is meaningful) is $\sim 0.1$~\AA. This result indicates that the state identified by classical MD corresponds to a metastable state also on the (more accurate) {\itshape ab-initio} potential, so confirming the reliability of our simulations. Finally, in order to take into account the anharmonicity of the potential, and the possibility that the metastable state identified by the classical simulation is not the most stable one, we ran a $5$~ps long {\itshape ab initio} MD simulation. In Tables \ref{tab:structuralData} and \ref{tab:structuralDataAbInitio} are reported the concentration and the structure of the N(Si$_n$N$_m$) units as obtained from the classical and {\itshape ab initio} MD simulations, respectively. By comparing the results reported in these two tables, we notice that the main difference among the classical and {\itshape ab initio} samples is the concentration of the N(Si$_4$) unit, which is lower than $1\%$ in the former and about $7\%$ in the latter. As for the bond length, the maximum difference between the two samples is $\sim 6 \%$.} 

\textcolor{black}{Analyzing more in detail about the structure of the N(Si$_n$N$_m$) units, we found that the N--N and Si--N bond length are rather unaffected by the composition of the unit as far as the coordination number of the central N atom is preserved (i.e. $m$ and $n$ at a fixed $n + m$), being N--N and N-Si bond length  $\sim 1.35$~\AA\ and $\sim 1.75$\AA, respectively.  Also the bond angle is essentially unaffected by the chemical composition of the N(Si$_n$N$_m$) units and its value is $\sim 120^\circ$. Such a bond angle indicates that the structure of these units is trigonal. These results are in agreement with experimental data.\cite{jncs-si3n4}
In conclusion, in the stoichiometric sample the N--N bonds are single bonds occurring in trigonal units at various concentration of Si and N. This result give evidence that the assumption that if two N atoms bond together they will form a N$_2$ molecule (that would leave the sample), which is the assumption made in the de Brito Mota et al. force field\cite{prb-68_205203} to make vanishing the N--N attractive interaction, is wrong. Below we shall show that this conclusion is independent on the stoichiometry of the sample.}

Before moving to the analysis of samples at different stoichiometries, we still have to evaluate the effect of the quenching rate on the concentration of Si and N coordination defects. This is a further validation of our computational setup and it is of great importance as the concentration of these defects is considered the structural features at the base of the interesting optical and electrical properties of this material. We do this by computing the percentage of Si$^n$ and N$^m$ in the sample, with $n = 3, 4, 5$ and  $m = 2, 3, 4$, where the superscript indicates the number of bonds formed by a given atom. The coordination number of a given atom is computed  according to the distance between this atom and its neighbors: two atoms are considered bonded if their distance is lower than the first minimum of the corresponding partial g(r).  In 
Fig.~5 
it is shown the percentage of Si$^n$ and N$^m$ as a function of the cooling rate. These data show that the concentration of N$^m$ is only marginally affected by the quenching rate. Essentially, N atoms are almost all 3-fold coordinated at any quenching rate apart for the case of the very high quenching rate dT/dt = $500$~K/ps and dT/dt = $100$~K/ps. The effect of the quenching rate on the coordination defects of Si is much more evident. At very high quenching rates the percentage of 5, 4 and 3-fold coordinated Si atoms is $\sim 22$~\%, $\sim 71$~\% and $\sim 7$~\%, respectively. The concentration of defects initially increases, then suddenly decreases with lower quenching rates and finally the sample converges to a composition of  \%Si$^3 \sim 24$~\%, \%Si$^4 \sim 75$~\% and \%Si$^5 \sim 1$~\%. It might appear surprising that the percentage at high and low quenching rates are close to each other. We believe that this is just due to chance.


Based on the analysis of the effect of the melting temperature and quenching rate on the stoichiometric sample, the non-stoichiometric samples were prepared with a melting temperature $T=6000$~K and a quenching rate $dT/dt = 5$~K/ps, that guarantees converged g(r) and composition of coordination defects. 
\textcolor{black}{As in the case of stoichiometric sample, we checked the reliability of our results by performing {\itshape ab-initio} geometry optimization and $5$~ps long MD simulations. Due to the high computational cost of {\itshape ab-initio} calculation on samples containing more than 200 atoms, we limited this analysis to $x = 0.4$ and $x = 1.8$ samples only, i.e. one in the sub and one in the supra-stoichiometric domain. Similarly to what found for the stoichiometric sample, the (averaged) classical and {\itshape ab-initio} configurations resulted to be very close to each other for both the sub and supra-stoichiometric samples (RMSD $\sim 0.2$ in both cases). This means that the state found by classical MD simulation is a metastable state also on the {\itshape ab initio} potential energy surface. We also compared classical and {\itshape ab initio} results at finite temperature (see Tables \ref{tab:structuralData} and \ref{tab:structuralDataAbInitio}). Both the concentration of the N(Si$_n$N$_m$) units and their structure as obtained from classical and {\itshape ab initio} are in agreement. 
This confirm that the results obtained from the classical simulations are reliable also in non-stoichiometric conditions. Finally, in order to further compare our results with experimental data, we also performed NPT MD simulations at room temperature and ambient pressure aimed at computing the density of all the samples. These results are reported in 
Fig.~6 
together with experimental data. The latter have been obtained by extrapolating experimental densities published in Ref. [\onlinecite{PhysRevB.42.5677}] to the stoichiometries studied in the present work. As a general remark, the computational trend is very similar to the experimental one, with the density growing with the concentration of N in the sample. More in detail, the computational values are very close to the experimental one, being the difference between $\sim 1.5$~\% and $\sim 5$~\% (see the inset of 
Fig.~6).}

Moving to the analysis of the structure of non stoichiometric samples, 
Fig.~7 
shows the total, and Si--Si and Si--N partial g(r) for SiN$_x$, with  x=0.4, 0.6, 0.8, 1.0, 1.2, 1.33 (Si$_3$N$_4$), 1.4 and 1.8. The N--N partial g(r) is not shown as, apart for the peak at $1.3$\AA, its shape is very broad and no insight on the structure of the samples can be obtained from it. In the following we shall analyze the peaks at $1.3$~\AA, $1.7$~\AA\ and the complex group of peaks between $2.2$ and $3$~\AA. As for the first peak, already identified as due to N--N pairs in the stoichiometric sample,  it increases with the nitrogen content. While this trend is in general not surprising, it is interesting to notice that this peak is already present in sub-stoichiometric conditions, starting from samples of composition $x = 1$. The main peak at $1.7$~\AA, due to Si--N pairs, initially increases with $x$ reaching a maximum at $x=1.33$ (stoichiometric composition) and then decreases. This is because at low $x$ there are not enough N to match with Si and the situation is reversed for $x > 1.33$. More interesting is the evolution of the complex series of peaks between $2.2$ and $3$~\AA. The g(r) in this interval is due to the superposition of the contributions of the three pairs N--N, Si--Si and Si--N. However, at low $x$ this peak is mainly due to Si--Si pairs, as it is confirmed by the comparison of the total and Si--Si partial g(r). With the increase of N content the shape of the g(r) in this range becomes more complex. At very high $x$ a satellite peak at $2.7$~\AA\ appears. This satellite peak is due to Si--N pairs, as can be seen by comparing the total and Si--N partial g(r). As for the dependency of the Si--Si partial g(r) with the composition, we notice that the peaks at $2.3$~\AA, due to chemically bonded Si atoms, decreases in intensity with the increase of $x$. At $x = 1.33$ this peak is very low, even though it never vanishes completely even at very high $x$. On the contrary, the intensity of the peak at $2.9$~\AA\ increases with $x$. This latter peak is due to non bonded Si--Si pairs of atoms which are both connected to the same N (see top-left panel of 
Fig.~7). 
The opposite trend of the two peaks can be explained by the transformation of Si(Si$_m$) units, which are present in sub-stoichiometric samples, into N(Si$_m$) units, typically present in stoichiometric and supra-stoichiometric systems. Very interesting is the analysis of the origin and the trend of the peak at $2.6$~\AA\ in the Si--N partial g(r). This peak is due to non-bonded Si--N pairs belonging to a unit in which the two atoms are both bonded to a N (see the pair connected by the dashed line in 
Fig.~3/B). 
It is worth to mention that this peak appears for the first time at $x =1.0$, in correspondence with the appearance of the peak at $1.3$~\AA, and then increases with $x$. On the basis of this analysis we conclude that in SiN$_x$ samples with $x \geq 1$ are presents units of the kind N(Si$_n$N$_m$), that is units in which a N is bonded to both Si and N  (see 
Fig.~3/A-C).
  \textcolor{black}{Similarly to the case of the stoichiometric sample, we analyzed the nature, concentration and structure of these units (see Table \ref{tab:structuralData}). In sub-stoichiometric samples we only found N(Si$_2$N) units. This is, of course, due to the low concentration of N. On the contrary, in the stoichiometric and in supra-stoichiometric samples we found N(Si$_2$N), N(SiN$_{2}$) and N(N$_{3}$) units (the latter was found only in the samples corresponding to $x = 1.6$ and $x = 1.8$). The concentration of N(Si$_n$N$_m$) units in general, and the concentration of  $m \geq 2$ units in particular, increases with $x$. The total concentration of these defective units is already very significant in the stoichiometric sample ($\sim 16.5\%$) and becomes higher than the concentration of N(Si$_3$) units at $x = 1.8$ ($58\%$ vs $39.5\%$). This latter result is confirmed by the {\itshape ab initio} MD simulations in which the concentration of N(Si$_n$N$_m$)) units resulted to be $\sim 53\%$, to be compared with a concentration of the N(Si$_3$) unit of $\sim 37\%$. Moving to the analysis of the structure of the N(Si$_n$N$_m$) units, both classical and {\itshape ab initio} simulations indicates that it is only marginally affected by the stoichiometry of the sample and the chemical composition of the unit. On the contrary, as expected, it strongly depends on the coordination of the central N atom. The average bond length in N(Si$_n$N$_m$) units in which the central N is three-fold coordinated (i.e. $n + m = 3$) is $d_{NN} = 1.33$~\AA\ and $d_{NSi} = 1.74$~\AA, and $d_{NN} = 1.42$~\AA\ and $d_{NSi} = 1.8$~\AA\ in classical and {\itshape ab initio} simulations, respectively. As for the (average) bond angle ($\alpha$), classical simulations give always an (almost perfect) trigonal structure for all the units in which the central N is three-fold coordinated ($\alpha \sim 119 ^\circ$). {\itshape Ab initio} simulation show a larger departure form this structure, with an (average) bond angle of $\sim 116 ^\circ$.}

Another very relevant structural parameter to consider is the concentration of coordination defects. In fact, as already mentioned in the introduction, both Si and N coordination detects are thought to be responsible for the electronic properties of a-SiN$_x$\cite{jr.:4190}. In 
Fig.~8 
we report the concentration of Si$^n$ (top) and N$^m$ (bottom) as a function of $x$. As for N, we did not observe any N$^2$ at any stoichiometry. On the contrary, there is a high concentration of N$^4$ defect (thought to be an electron trapping defect)  at low $x$. However, the concentration of this type of defect decreases with $x$ and becomes negligible for $x = 1.2$. The situation is different for Si. At $x = 0.4$ the concentration of Si$^3$, Si$^4$ and  Si$^5$ is $15$~\%, $79$~\% and $6$~\%, respectively. By comparison with the concentration of similar defects in amorphous Si (a-Si), we conclude that the effect of low concentration of N is to increase the number of under-coordinated Si. 
In fact, in  a-Si, by using the same potential and amorphization protocol, we have measured 3\% of Si$^3$, 93\% of Si$^4$ and 4\% of Si$^5$  (these values are in line with ab-initio MD simulation results\cite{PhysRevB.44.11092}). 
At variance with N, the concentration of the Si coordination defects remain large at any N content. However, starting from the stoichiometric composition we observe a reduction of their concentration. Much the same as for N, most of the Si coordination defects are due to under-coordination. However, at very low and very high N content there is also a non negligible amount of $5$-fold coordinated Si atoms.

\section{Conclusions
\label{sec:Conclusions}
}
We studied the structural features and the coordination defects of silicon nitride samples at various compositions as obtained from the quench-from-the-melt method. Our results are in good agreement with X-ray and neutron diffraction data\cite{jncs-si3n4-xray,jncs-si3n4}. In particular, our simulations show that the peak at $1.3$~\AA\ in the experimental g(r) is due to the formation of N(Si$_n$N$_m$) units. This finding is consistent with other experimental data, namely
Electron Spin Resonance (ESR) measures \cite{jjap.32.876},  and X-ray photoemission spectroscopy, electron-energy-loss spectroscopy and optical absorption\cite{PhysRevB.42.5677} experiments on hydrogenated and non-hydrogenated SiN$_x$ samples. 
Also the density at room temperature and ambient pressure obtained from NPT-MD simulations is in good agreement with experimental results.\cite{PhysRevB.42.5677} 
This is a further confirm that the Billeter et al. force field \cite{PhysRevB.73.155329, PhysRevB.79.169904} is adequate for simulating semiconductors of SiN$_x$ type. 
We believe that the inability of previous computational works to reproduce the features mentioned above was due to either the force field used for the simulations, which explicitly prevented the formation of N--N bonds, or to the quenching-from-the-melt parameters used in the simulations. In particular, we think that in previous simulations\cite{lin-optoAndAdvMat-4_543,lin-optmat-93-93, kroll2001,prb-68_205203, PhysRevB.54.R15594, PhysRevB.41.12197,PhysRevB.65.073202,PhysRevB.58.8323,prb80-144201,prb82-205201} it was used a too low melting temperatures and/or a too short melting time. Since the breaking (of Si--N) and the formation of (N--N) bonds is a rare events, i.e. it occurs on a time scale longer than the typical duration of an MD simulation, the conditions used in previous simulations do not allow them to occur on the time scale of simulations. Other structural features, and as a consequence electronic structural features, might be also affected by the inadequacy of the force field or the amorphization protocol.

We also analyzed the effect of the quenching rate on the structure of the sample, concluding that very low quenching rates are needed to obtain well converged amorphous samples.

We also studied the structural features of non-stoichiometric SiN$_x$ samples, which are of great technological interest. We found that at low $x$ the sample contains Si(N$_m$Si$_{4-m}$) and NSi$_3$ units. However, already at $x = 1$, i.e. still in sub-stoichiometric conditions, other units containing N--N bonds are formed. These units are not N$_2$ molecules, rather N(Si$_n$N$_m$) units in which a nitrogen atom is bonded to both Si and N atoms.

Finally, we studied the concentration of coordination defects as a function of the stoichiometry of the sample. These defects are considered relevant for the electronic properties of this material, especially for its use in the non-volatile memory device field. We found that at very low N concentration there is a significant amount of 4-fold coordinated N atoms while the concentration of N$^2$ defects is negligible at all $x$. Defects of this type are thought to be responsible for the electron trapping properties of SiN$_x$. As for Si, we found that there is a significant amount of coordination defects at any N content. In particular, these defects are mostly associated to under-coordinated Si, apart at very low and very high $x$ where there is also a non negligible amount of Si$^5$.

\acknowledgements
The authors wish to acknowledge the SFI/HEA Irish Centre for High-End Computing (ICHEC) for the provision of computational facilities. One of the authors (S. M.) acknowledges SFI Grant 08-IN.1-I1869 for the financial support.

\section*{Figure captions}
{Fig. 1: Pair correlation function of liquid stoichiometric samples at various melting temperatures after $100$~ps of simulation. In the inset are displayed the partial N--N g(r) at the corresponding melting temperatures.}
\\ \\

{Fig. 2: Pair correlation function of stoichiometric amorphous samples obtained at various cooling rates. In the inset is shown the relative concentration of Si--Si pairs with respect to Si--N pairs as a function of the cooling rate.}
\\ \\

{Fig. 3: (color online) Snapshots of the N(Si$_2$N) (A), N(SiN$_2$) (B) and N(N$_3$) (C) units. Circles indicates the units containing N--N bonds; N and Si atoms part of these units are bigger and colored in light gray and Orange, respectively.  }
\\ \\

{Fig. 4: Radial Distribution Function (RDF) of the stoichiometric sample obtained from a quenching-from-the-melt run at $T_m = 6000$~K and $dT/dt = 5$~K/ps. In the inset is reported the experimental RDF obtained from neutron diffraction. data\cite{jncs-si3n4}}
\\ \\

{Fig. 5: Si$^n$ (top) and N$^m$ (bottom) concentration of coordination defect as a function of the cooling rate.}
\\ \\

{Fig. 6: Computation vs experimental density at various stoichiometries. Experimental data have been extrapolated from data published in Ref. [\onlinecite{PhysRevB.42.5677}].}
\\ \\

{Fig. 7: Total (bottom), Si--Si partial (top left) and N--N partial (top right) pair correlation function of samples at various stoichiometries.}
\\ \\

{Fig. 8: Concentration of Si$^n$ (top) and N$^m$ (bottom) species as a function of the stoichiometry of the sample.}

\newpage

\begin{table}
\caption{List of simulations and corresponding quenching-from-the-melt parameters. Each simulation was followed by a $100$~ps of thermalization at $300$~K and a $10$~ps simulation over which the observables have been computed.\label{tab:amorphizationProtocol}}
\begin{tabular}{c||cc|c}
Name               & \multicolumn{2}{c|}{Melting} & \multicolumn{1}{c}{Cooling} \\ 
                         & T$_m$ [K]  & t$_m$ [ps] & dT/dt [K/ps] \\ \hline
5000-100-5       &   5000         &     100        &     5  \\ \hline \hline
5800-100-5       &   5800         &     100        &     5  \\ \hline \hline
6000-100-1       &   6000         &     100        &     1  \\ \hline 
6000-100-5       &   6000         &     100        &     5  \\ \hline 
6000-100-10     &   6000         &     100        &     10  \\ \hline 
6000-100-50     &   6000         &     100        &     50  \\ \hline
6000-100-100   &   6000         &     100        &     100  \\ \hline 
6000-100-500   &   6000         &     100        &     500  \\ \hline 
7000-100-5       &   7000         &     100        &     5  \\ 
\end{tabular}
\end{table}%

\begin{table}
\caption{Density used for the NVT simulations as a function of the composition of the sample.}
\label{tab:density}
\begin{center}
\begin{tabular}{c|c|c}

sample & $x$     &  density (g/cm$^3$) \\ \hline

Si$_{64}$N$_{160}$  & $0.400$ & $2.643$ \\
Si$_{84}$N$_{140}$  & $0.600$ & $2.771$ \\
Si$_{100}$N$_{124}$ & $0.806$ & $2.890$ \\
Si$_{112}$N$_{112}$ & $1.000$ & $2.991$ \\
Si$_{101}$N$_{123}$ & $1.218$ & $3.091$ \\
Si$_{96}$N$_{128}$  & $1.333$ & $3.140$ \\
Si$_{93}$N$_{131}$  & $1.409$ & $3.169$ \\
Si$_{86}$N$_{138}$  & $1.605$ & $3.237$ \\
Si$_{89}$N$_{144}$  & $1.800$ & $3.294$ \\

\end{tabular}
\end{center}
\end{table}

\newpage

\begin{table}
\caption{Concentration and structural data of the most abundant N(Si$_n$N$_m$) and Si(Si$_n$N$_m$) units in the melted Si$_3$N$_4$ sample at various T. The structural data reported are the N--N (dNN) and N--Si (dNSi) bond length, and the average bond angle ($\alpha$). These data are computed by  time averaging the corresponding instantaneous quantities along the classical and {\itshape ab initio} MD simulations. Distances are reported in \AA\ and angles in $^\circ$.\label{tab:structuralDataMelt}} 
\begin{tabular}{cc|c|c|c||cc|c|c|c}
 \multicolumn{5}{c||}{Classical Simulations} & \multicolumn{5}{c}{{\itshape Ab initio} simulations} \\ \hline
 \multicolumn{2}{c|}{} & $5800$~K & $6000$~K & $7000$~K & \multicolumn{2}{c|}{} &$5800$~K & $6000$~K & $7000$~K \\ \hline
 
 & \% & 58   & 48 & 41 &   & \% & 25  & 20 & 15\\[-2ex]
N(Si$_3$) & d$_{NSi}$ & 1.79 & 1.80 & 1.81 &  Si(N$_4$)   & d$_{SiN}$ & 1.78 & 1.78 & 1.78  \\ [-2ex]
 &  $\alpha$ & 108 & 116  & 116 &       &  $\alpha$  & 108 & 107 & 107 \\ [-1ex] \hline 
                                          
 & \% & 23 & 23 & 22 &   & \% & 21 & 21 & 19 \\ [-2ex]
N(Si$_2$N) & d$_{NN}$ & 1.41  & 1.42 & 1.43  & Si(N$_3$Si)  & d$_{SiN}$ & 1.77  & 1.77 & 1.77  \\ [-2ex]
 & d$_{NSi}$ & 1.81 & 1.81 & 1.81  &  & d$_{SiSi}$ & 2.44 & 2.44 & 1.43 \\ [-2ex]
 &  $\alpha$       & 116 & 116 & 116 &    &  $\alpha$       & 106 & 106 & 105  \\ [-1ex] \hline        
                                                         
 & \% & 13 & 14.5 & 18.5 & & \% & 20.5 & 19 & 16 \\ [-2ex]
 N(Si$_2$) & d$_{NSi}$ & 1.77  & 1.76 & 1.76  & Si(N$_3$)  & d$_{SiN}$ & 1.77  & 1.76 & 1.77 \\ [-2ex]
 &  $\alpha$ & 121 & 121 & 121 &   &  $\alpha$       & 112 & 112 & 111  \\ [-1ex] \hline 
                                          
 & \% & 3.5 & 4.5 & 6.5 & & \% & 8.5 & 10 & 12 \\[-2ex]
N(SiN) & d$_{NN}$ & 1.41 & 1.38 & 1.40 & Si(SiN$_2$) & d$_{SiN}$ & 1.77 & 1.76 & 1.76 \\ [-2ex]
 & d$_{NSi}$ & 1.77 & 1.77 & 1.78 &   & d$_{SiSi}$ & 2.40 & 2.39 & 2.39 \\ [-2ex]
 & $\alpha$ & 122 & 121 & 121 &  &  $\alpha$ & 110 & 109 & 109  \\  [-1ex]\hline      

   & \% & 3.5 & 4 &  4.5 & & \% & 6 & 7 &  6 \\[-2ex]
N(SiN$_2$) & d$_{NN}$ & 1.41 & 1.41 & 1.43 & Si(SiN$_4$) & d$_{SiN}$ & 1.78 & 1.78 & 1.78 \\ [-2ex]
 & d$_{NSi}$ & 1.82 & 1.82 & 1.82 &  & d$_{SiSi}$ & 2.53 & 2.53 & 1.52 \\ [-2ex]
 & $\alpha$ & 116 & 116 & 116 &  & $\alpha$ & 103 & 102 & 102 \\ [-1ex]\hline
                                          
 & \% & 1.5 & 2 &  2 & & \% & 5 & 6 &  8 \\[-2ex]
N(Si$_4$) & d$_{NSi}$ & 1.90 & 1.90 & 1.90 & Si(Si$_2$N$_2$) & d$_{SiN}$ & 1.77 & 1.76 & 1.76 \\ [-2ex]
 & $\alpha$ & 108 & 108 & 108 &  & d$_{SiSi}$ & 2.42 & 2.42 & 2.41 \\ [-2ex]
 &  &  &  &  &  & $\alpha$ & 105 & 105 & 104 \\ [-1ex]\hline 
   
 & \% & 1 & 1.5 & 2 &  & \% & 4 & 5 & 6 \\[-2ex]
N(Si$_3$N) & d$_{NN}$ & 1.51 & 1.52 & 1.53 & Si(Si$_2$N$_3$)  & d$_{SiN}$ & 1.76 & 1.76 & 1.76 \\ [-2ex]
 & d$_{NSi}$ & 1.90 & 1.90 & 1.90 &  & d$_{SiSi}$ & 2.49 & 2.49 & 2.47 \\ [-2ex]
 & $\alpha$ & 108 & 107 &  107 &  &  $\alpha$ & 102 & 102 &  101 \\ [-1ex]\hline 
                        
\multicolumn{5}{c||}{}  & & \% & 3.5 & 4 & 4 \\[-2ex]
\multicolumn{5}{c||}{}  &  Si(N$_2$)  & d$_{SiN}$ & 1.77 & 1.76 & 1.76 \\[-2ex] 
\multicolumn{5}{c||}{}  &    &  $\alpha$   & 116 & 116 & 115 \\ [-1ex] \hline \hline

\end{tabular}
\end{table}

\newpage

\begin{table} 
\caption{Concentration and structural data of the most abundant N(Si$_n$N$_m$) units ( $\% \geq 1$) in the samples at various composition.The structural data reported are the N--N (dNN) and N--Si (dNSi) bond length, and the average bond angle ($\alpha$). These data are computed by  time averaging the corresponding instantaneous quantities along the classical and {\itshape ab initio} MD simulations. Distances are reported in \AA\ and angles in $^\circ$.\label{tab:structuralData}} 
\begin{tabular}{c|c|c||c|c|c||c|c|c||c|c|c||c|c|c}
\multicolumn{3}{c||}{x = 0.4} & \multicolumn{3}{c||}{x = 0.6} & \multicolumn{3}{c||}{x = 0.8} & \multicolumn{3}{c||}{x = 1.0} & \multicolumn{3}{c}{x = 1.2} \\ [-1ex]\hline 
& \% & 75 & & \% & 89 & & \% & 95 & & \% & 88 &  & \% & 86 \\[-2ex]
   N(Si$_3$)                  & d$_{NSi}$ & 1.71 &  N(Si$_3$) & d$_{NSi}$ & 1.71 & N(Si$_3$) & d$_{NSi}$ & 1.71 & N(Si$_3$) & d$_{NSi}$ & 1.71 & N(Si$_3$)  & d$_{NSi}$ & 1.72 \\ [-2ex]
                                          &  $\alpha$       & 116 &    &  $\alpha$   & 119 &   &  $\alpha$  & 119 &    &  $\alpha$   & 119 &    &  $\alpha$  & 118  \\ [-1ex]\hline 

 & \% & 25 &  & \% & 11 & & \% & 5 &  & \% & 6 &   & \% & 13 \\[-2ex]
   N(Si$_4$)             & d$_{NSi}$ & 1.79 & N(Si$_4$)  & d$_{NSi}$ & 1.79 &  N(Si$_4$)  & d$_{NSi}$ & 1.78 &  N(Si$_4$) & d$_{NSi}$ & 1.79 & N(Si$_2$N)  & d$_{NSi}$ & 1.72 \\ [-2ex]
                                          &  $\alpha$   & 109 &   &  $\alpha$  & 109 &    &  $\alpha$   & 109  &   &  $\alpha$   & 109 &   & d$_{NN}$ & 1.32  \\  [-2ex]
                                          &                   &        &   &                  &        &    &                   &         &   &                   &        &   & $\alpha$  & 118  \\  [-1ex]\hline

\multicolumn{3}{c}{}  & \multicolumn{3}{c}{}  & \multicolumn{3}{c||}{} &   & \% & 6 & \multicolumn{3}{c}{}   \\[-2ex]
\multicolumn{3}{c}{}  & \multicolumn{3}{c}{}  & \multicolumn{3}{c||}{} & N(Si$_2$N) &  d$_{NSi}$ & 1.72 & \multicolumn{3}{c}{}   \\ [-2ex]
\multicolumn{3}{c}{}  & \multicolumn{3}{c}{}  & \multicolumn{3}{c||}{} &  &  d$_{NN}$ & 1.32 & \multicolumn{3}{c}{}    \\  [-2ex]
\multicolumn{3}{c}{}  & \multicolumn{3}{c}{}  & \multicolumn{3}{c||}{} &  &  $\alpha$  & 119  & \multicolumn{3}{c}{}   \\ [-0.5ex] \hline \hline

\multicolumn{3}{c||}{Si$_3$N$_4$} & \multicolumn{3}{c||}{x = 1.4} & \multicolumn{3}{c||}{x = 1.6} & \multicolumn{3}{c||}{x = 1.8} & \multicolumn{3}{c}{} \\[-1ex] \cline{1-12} 
 & \% & 81 &  & \% & 67 & & \% & 53.5 & & \% & 41.5 & \multicolumn{3}{c}{}  \\[-2ex]
         N(Si$_3$)               & d$_{NSi}$ & 1.73  & N(Si$_3$)  & d$_{NSi}$ & 1.72  & N(Si$_3$)   & d$_{NSi}$ & 1.73 & N(Si$_2$N)  & d$_{NSi}$ & 1.74 & \multicolumn{3}{c}{} \\ [-2ex]
                                          &  $\alpha$  & 119   &      &  $\alpha$  & 119 &     &  $\alpha$   & 119 &  & d$_{NN}$ & 1.32  & \multicolumn{3}{c}{}  \\   [-1ex]                                         & & &  & & &  & & &  &  $\alpha$  & 119  & \multicolumn{3}{c}{}  \\ [-1ex]\cline{1-12}  
                                          
& \% & 15 & & \% & 25 &  & \% & 32.5 & & \% & 39.5 &  \multicolumn{3}{c}{}  \\[-2ex]
     N(Si$_2$N)              & d$_{NSi}$ & 1.75 & N(Si$_2$N)    & d$_{NSi}$ & 1.73       &  N(Si$_2$N)  & d$_{NSi}$ & 1.74 &  N(Si$_3$)  & d$_{NSi}$ & 1.73  & \multicolumn{3}{c}{}  \\ [-2ex]
                                           & d$_{NN}$ & 1.34 &     & d$_{NN}$ & 1.33 &      & d$_{NN}$ & 1.33 &     &  $\alpha$  & 119  & \multicolumn{3}{c}{}  \\  [-2ex]
                                          &  $\alpha$  & 119 &     &  $\alpha$  & 119  &     &  $\alpha$  & 119 & & & & \multicolumn{3}{c}{} \\[-1ex] \cline{1-12}

 & \% & 1.5 &  & \% & 7  &  & \% & 9.5 & & \% & 14.5 & \multicolumn{3}{c}{}   \\[-2ex]
     N(SiN$_2$)                  & d$_{NSi}$ & 1.76 & N(SiN$_2$)  & d$_{NSi}$ & 1.73  &     N(SiN$_2$)      & d$_{NSi}$ & 1.75 & N(SiN$_2$)  & d$_{NSi}$ & 1.76 & \multicolumn{3}{c}{}   \\ [-2ex]
                                          & d$_{NN}$ & 1.33 &   & d$_{NN}$ & 1.32  &  & d$_{NN}$ & 1.32  &   & d$_{NN}$ & 1.33 & \multicolumn{3}{c}{}    \\  [-2ex]
                                          &  $\alpha$  & 119  &   &  $\alpha$  & 120  &   &  $\alpha$  & 119 &    &  $\alpha$  & 118   & \multicolumn{3}{c}{}   \\  [-1ex]\cline{1-12} 

\multicolumn{3}{c||}{} & \multicolumn{3}{c||}{}  &  & \% & 1.5 & & \% & 3  & \multicolumn{3}{c}{}   \\[-2ex]
                            \multicolumn{3}{c||}{} & \multicolumn{3}{c||}{} & N(N$_3$) & d$_{NN}$ & 1.32  &  N(N$_3$)  & d$_{NN}$ & 1.33 & \multicolumn{3}{c}{}    \\  [-2ex]
                            \multicolumn{3}{c||}{} & \multicolumn{3}{c||}{}   &   &  $\alpha$  & 118 &    &  $\alpha$  & 119   & \multicolumn{3}{c}{}   \\ [-1ex]  \cline{1-12} 
 
\multicolumn{3}{c||}{} & \multicolumn{3}{c||}{}  & & \% & 1.5 & \multicolumn{3}{c||}{} & \multicolumn{3}{c}{}   \\[-2ex]
                            \multicolumn{3}{c||}{} & \multicolumn{3}{c||}{} & N(Si$_3$N) & d$_{NSi}$ & 1.86   & \multicolumn{3}{c||}{} &\multicolumn{3}{c}{}    \\  [-2ex]
                            \multicolumn{3}{c||}{} & \multicolumn{3}{c||}{} &   & d$_{NN}$ & 1.37  & \multicolumn{3}{c||}{} &\multicolumn{3}{c}{}    \\  [-2ex]
                            \multicolumn{3}{c||}{} & \multicolumn{3}{c||}{}   &     &  $\alpha$  & 109  & \multicolumn{3}{c||}{} & \multicolumn{3}{c}{}   \\ [-1ex] \cline{1-12}  \cline{1-12} 

\end{tabular}
\end{table}

\newpage

\begin{table}
\caption{Same data as in Table \ref{tab:structuralData} for samples at selected stoichiometries ($x = 0.4$, Si$_3$N$_4$ and $x =1.8$) as obtained form $5$~ps long {\itshape ab initio} MD simulations.}
\label{tab:structuralDataAbInitio}
\begin{center}
\begin{tabular}{c|c|c||c|c|c||c|c|c}
\multicolumn{3}{c||}{x = 0.4} & \multicolumn{3}{c||}{Si$_3$N$_4$} & \multicolumn{3}{c}{x = 1.6}  \\ \hline 
\multirow{4}{*}{N(Si$_3$)} & \% & 77.5 & \multirow{4}{*}{N(Si$_3$)} & \% & 73.5 & \multirow{4}{*}{N(Si$_2$N)} & \% & 38.5 \\
                                          & d$_{NSi}$ & 1.82 &    & d$_{NSi}$ & 1.79 &   & d$_{NSi}$ & 1.79   \\ 
                                         &  $\alpha$  & 117 &      &  $\alpha$  & 116  &   & d$_{NN}$ & 1.42  \\ 
                                  & & &  & &   & &  $\alpha$  & 115  \\  \hline

\multirow{4}{*}{N(Si$_4$)} & \% & 22.5 & \multirow{4}{*}{N(Si$_2$N)} & \% &14.5        & \multirow{4}{*}{N(Si$_3$)} & \% & 37\\
                                          & d$_{NSi}$ & 1.94 &   & d$_{NSi}$ & 1.80 &   & d$_{NSi}$ & 1.77    \\ 
                                         &  $\alpha$   & 110 &      & d$_{NN}$ & 1.45 &    &  $\alpha$  & 116    \\ 
                                         &     &  &      & $\alpha$ & 116 &    &   &     \\  \hline 

\multicolumn{3}{c||}{} & \multirow{4}{*}{N(SiN$_2$)} & \% & 1.5 & \multirow{4}{*}{N(SiN$_2$)} & \% & 12.5  \\
\multicolumn{3}{c||}{} &    & d$_{NSi}$ & 1.83&   & d$_{NSi}$ & 1.83    \\ 
\multicolumn{3}{c||}{} &    & d$_{NN}$ & 1.44 &   & d$_{NN}$ &  1.41    \\ 
\multicolumn{3}{c||}{} &   & $\alpha$ & 115  &  & $\alpha$  &  114    \\  \hline 

\multicolumn{3}{c||}{} & \multicolumn{3}{c||}{} & \multirow{3}{*}{N(N$_3$)} & \% &  2  \\
\multicolumn{3}{c||}{} &  \multicolumn{3}{c||}{} &  & d$_{NN}$ & 1.42    \\ 
\multicolumn{3}{c||}{} &  \multicolumn{3}{c||}{} &  & $\alpha$  &  111    \\  \hline 

\end{tabular}
\end{center}
\end{table}

\begin{figure}[h]
\includegraphics[width=0.9\textheight,angle=270]{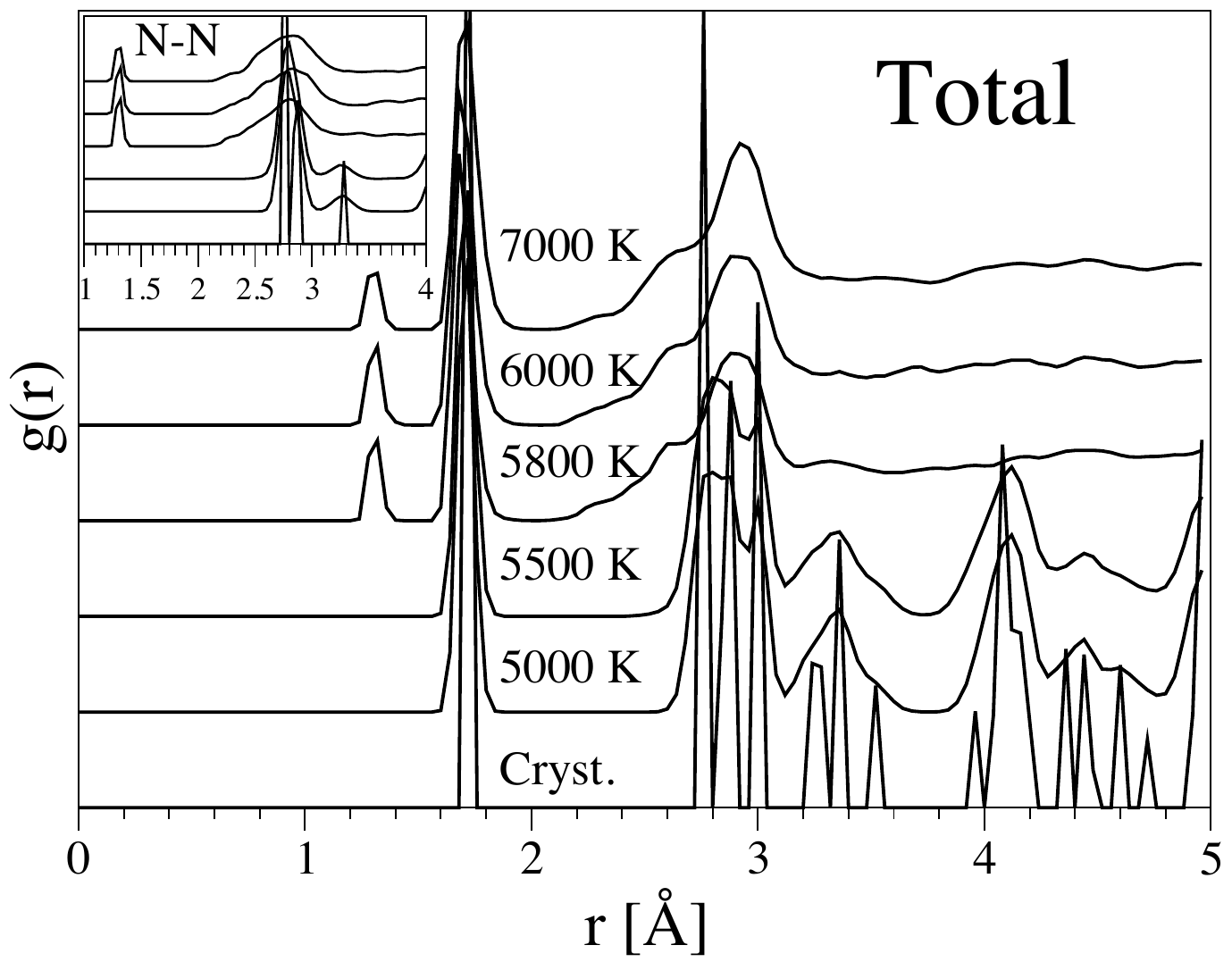}
\caption{\label{Fig:gr-melt}} 
\end{figure}

\begin{figure}[h]
\includegraphics[width=0.9\textheight,angle=270]{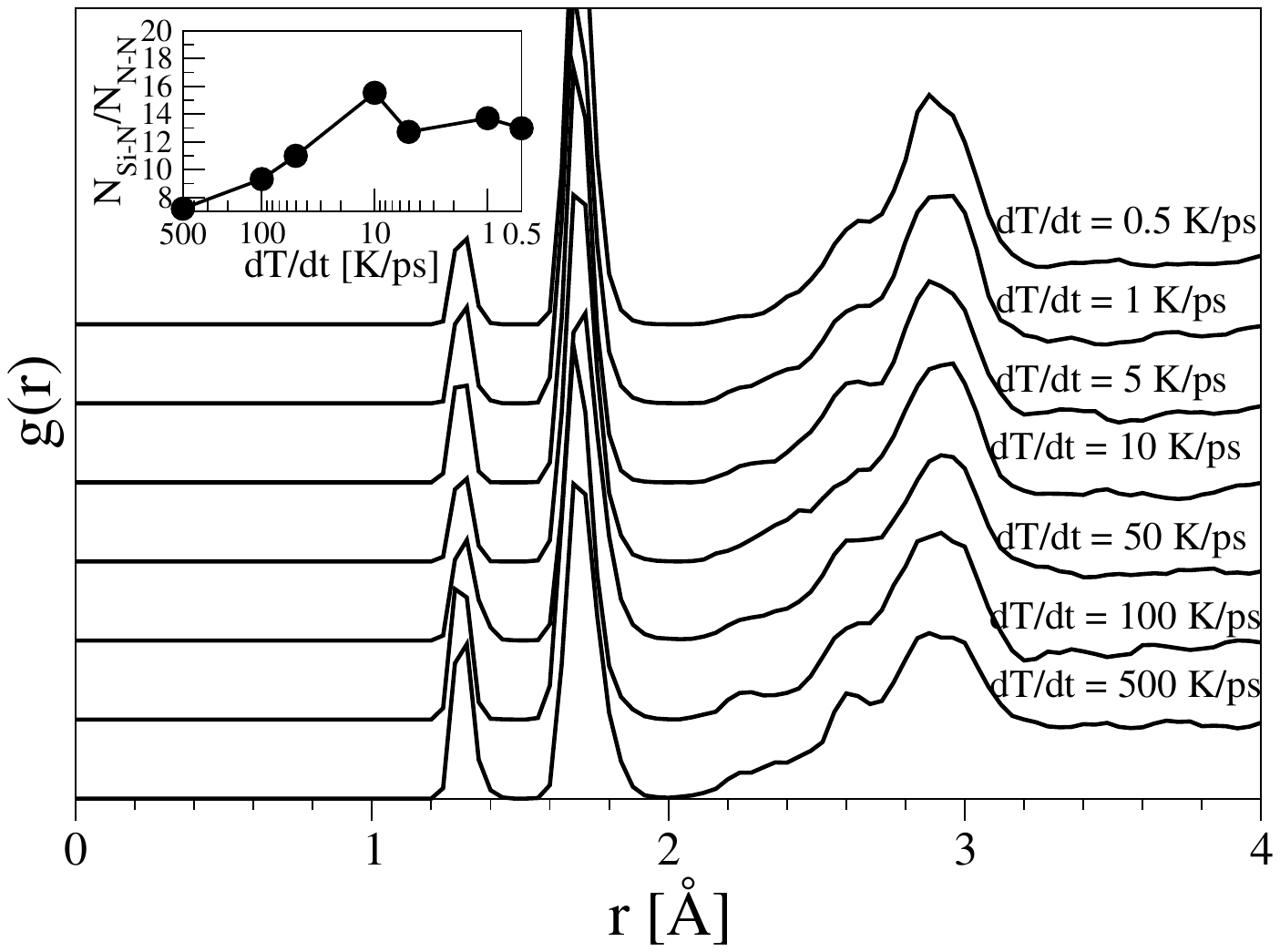}
\caption{\label{Fig:gr-cooling}}
\end{figure}

\begin{figure}[h]
\includegraphics[width=0.8\textheight,angle=270]{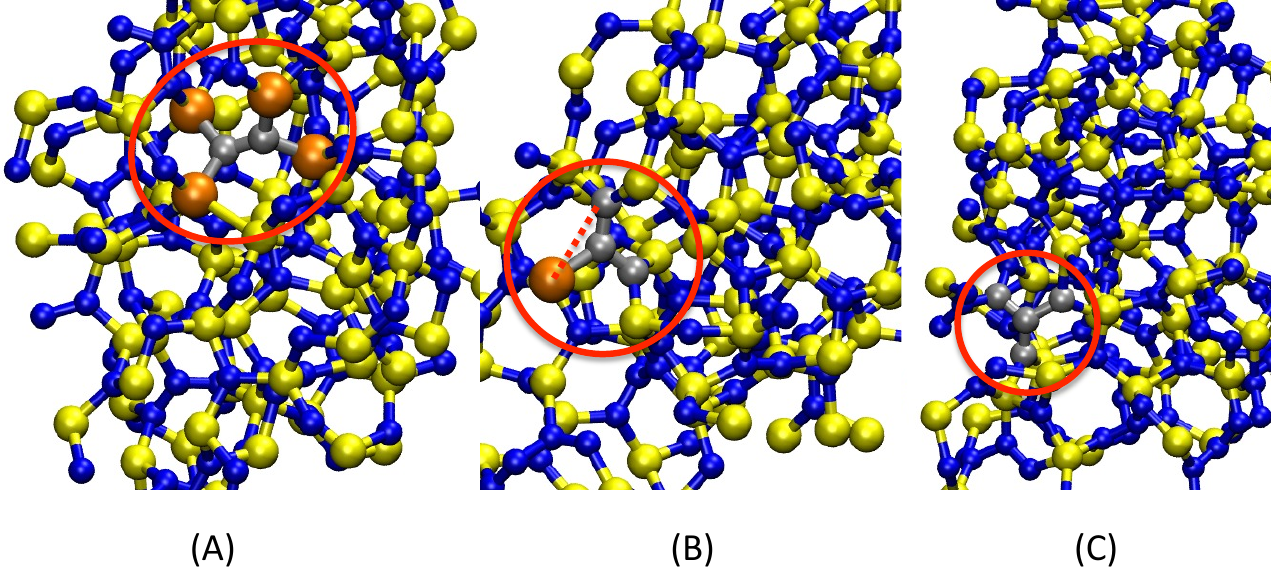}
\caption{\label{Fig:N-NmSi3-m:structures}}
\end{figure}

\begin{figure}[h]
\includegraphics[width=0.9\textheight,angle=270]{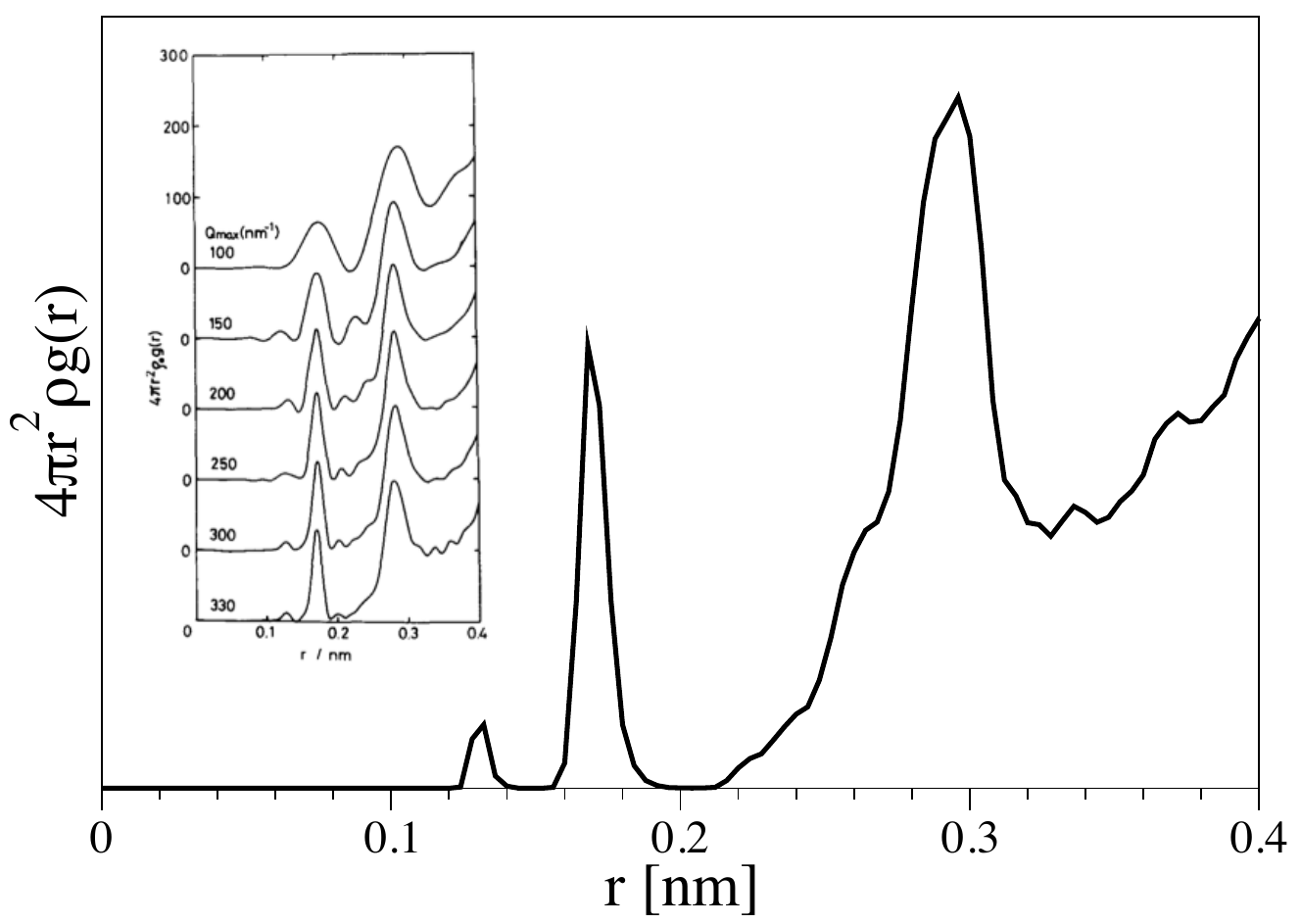}
\caption{\label{Fig:rdf}}
\end{figure}

\begin{figure}[h]
\includegraphics[width=0.9\textheight,angle=270]{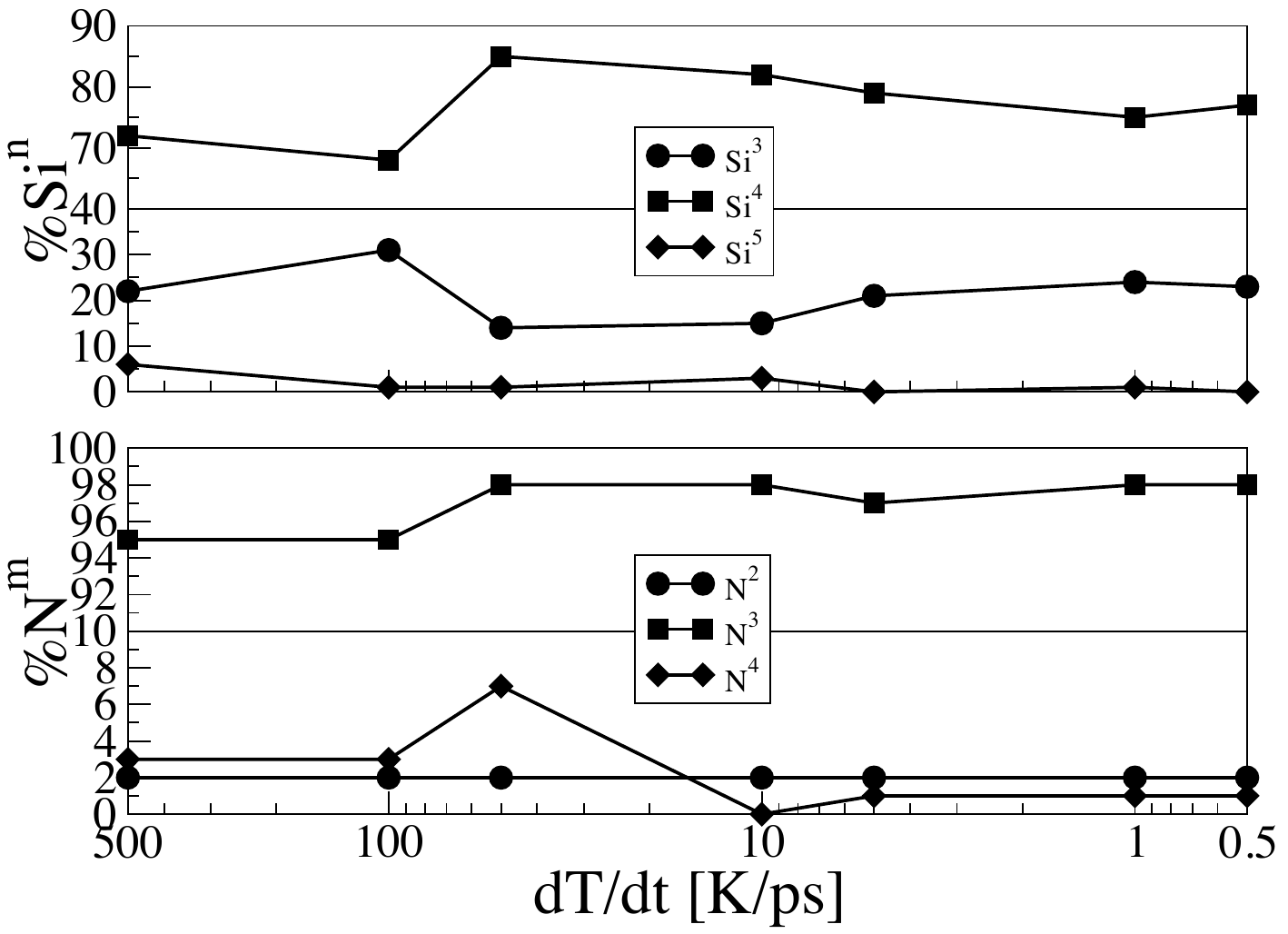}
\caption{\label{Fig:CoordCoolingRate}}
\end{figure}

\begin{figure}[h]
\includegraphics[width=0.9\textheight,angle=270]{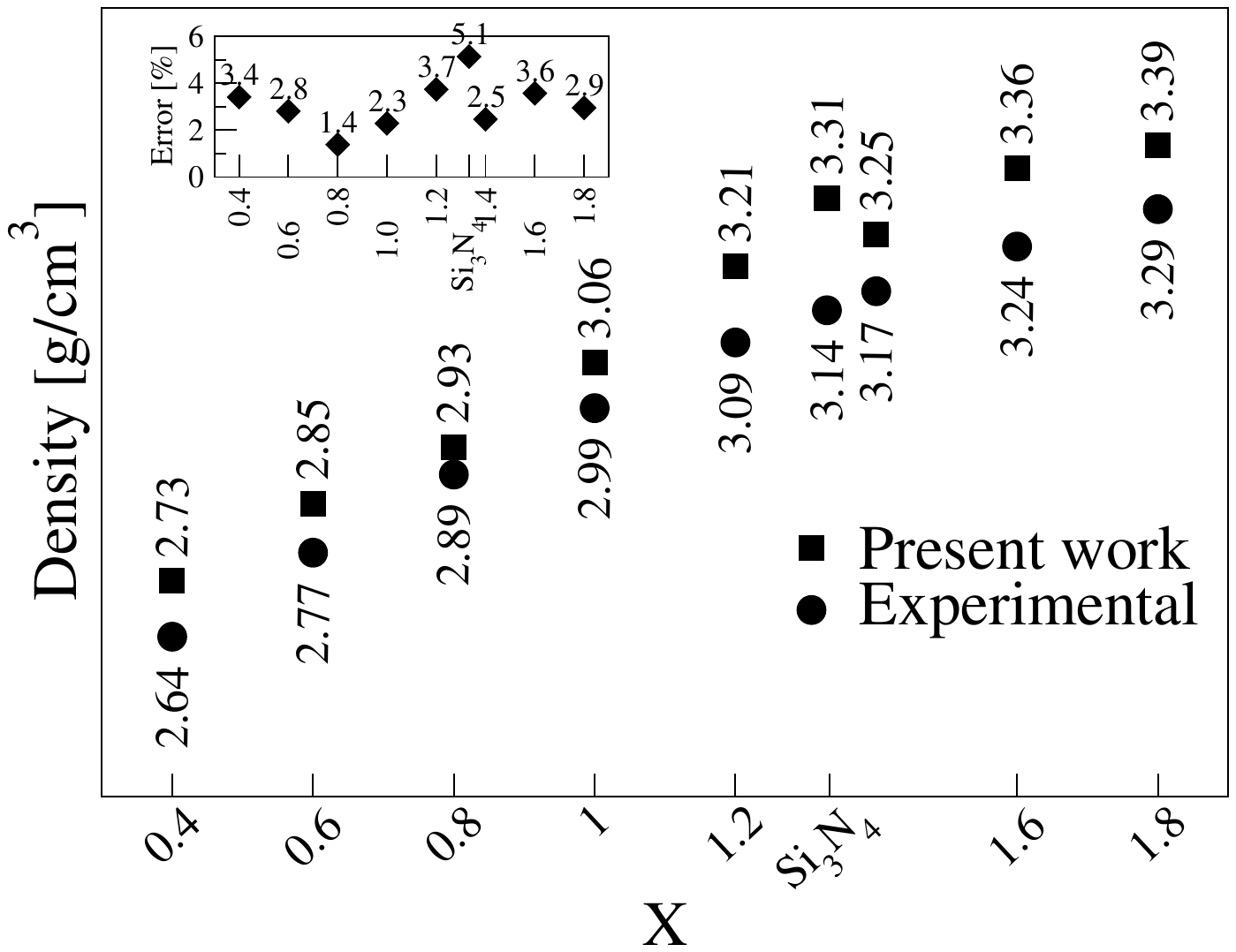}
\caption{\label{Fig:density}}
\end{figure}

\begin{figure}[h]
\includegraphics[width=0.9\textheight,angle=270]{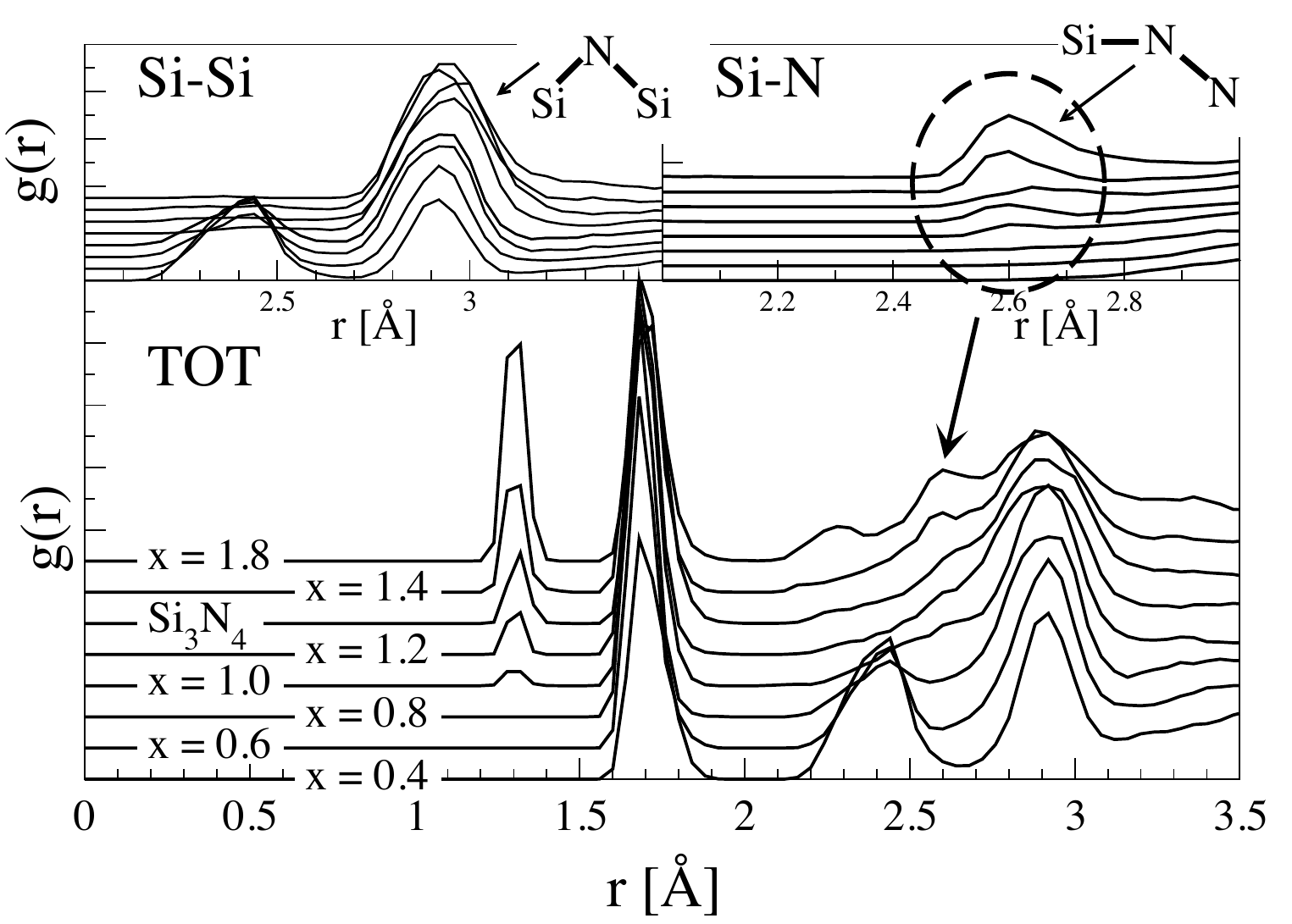}
\caption{\label{Fig:gr-variousX-partial}}
\end{figure}

\begin{figure}[h]
\includegraphics[width=\textheight,angle=270]{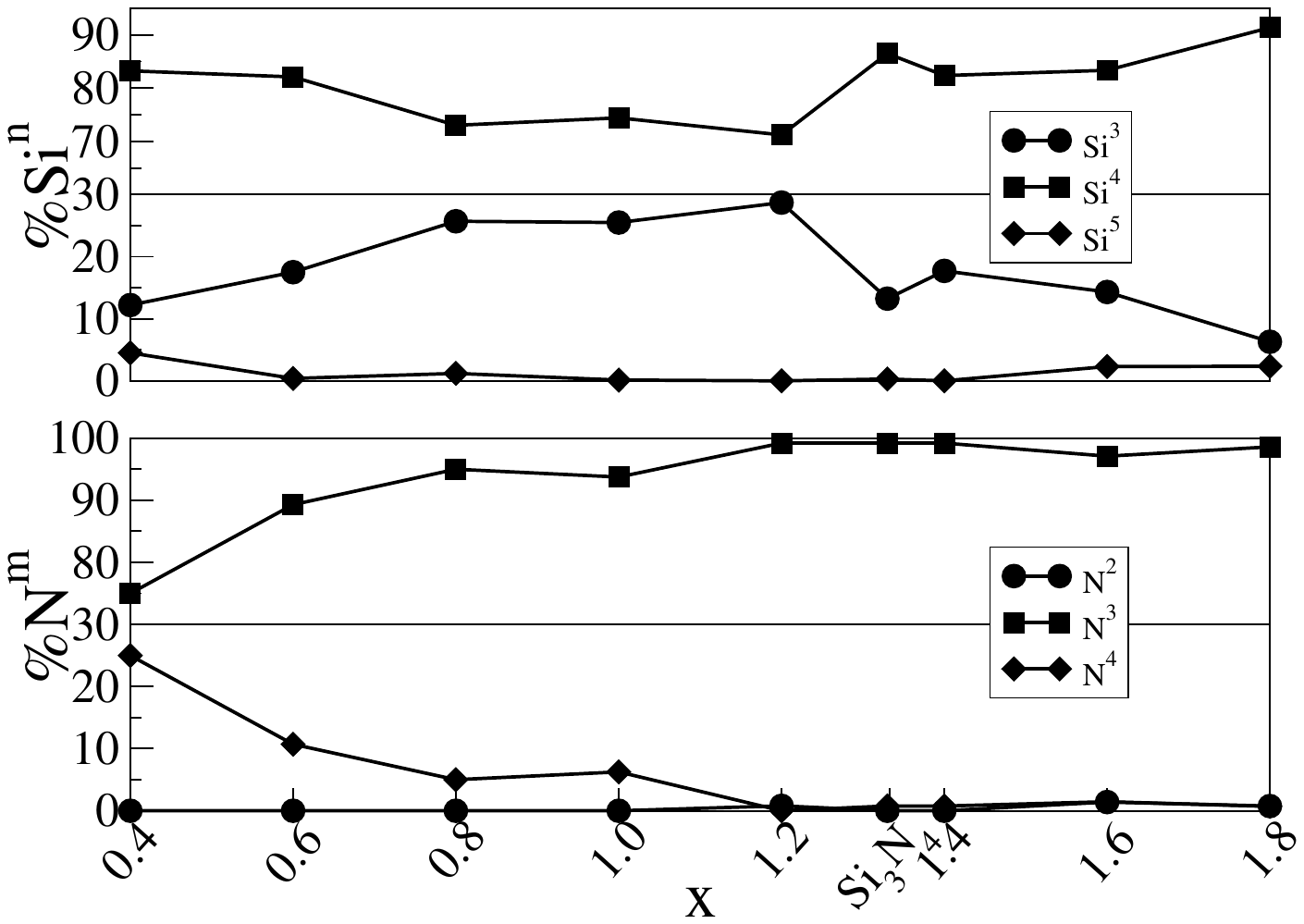}
\caption{\label{Fig:defects-variousX} }
\end{figure}

\end{document}